# On the rheology of red blood cell suspensions with different amounts of dextran: separating the effect of aggregation and increase in viscosity of the suspending phase

Daniel Flormann[1], Katharina Schirra[2], Thomas Podgorski[3] and Christian Wagner[1]

[1]Experimental physics, Saarland University, 66123 Saarbrücken, Germany

[2]Winterberg hospital, Saarland, 66119 Saarbrücken, Germany

[3]Laboratoire Interdisciplinaire de Physique, CNRS-UMR 5588, Université Grenoble, B.P. 87, 38402 Saint Martin d'Hères Cedex, France

## Abstract

We investigate the shear thinning of red blood cell - dextran suspensions. Microscopic images show that at low polymer concentration, aggregation increases with increasing concentration until it reaches a maximum and then decreases again to non-aggregation. This bell shape dependency is also deduced from the rheological measurements, if the data are correctly normalized by the viscosity of the suspending phase since a significant amount of polymers adsorb to the cell surfaces. We find that the position of the maximum of this shear rate dependent bell shape increases with increasing viscosity of the suspending phase, which indicates a that the dynamic process of aggregation and disaggregation is coupled via hydrodynamic interactions. This hydrodynamic coupling can be suppressed by characterizing a suspension of 80% hematrocrit which yields good agreement with the results from the microscopical images.

## Keywords

Blood, suspension, shear thinning, dextran

## Introduction

It is known since the seminal works by Merril et al. (Merill et al. 1966) and Chien et al. (Chien et al. 1967) that the shear thinning of blood is caused by a reversible aggregation-dissociation mechanism induced by the plasma proteins, mainly the fibrinogen (Baskurt et al. 2012). The viscosity curve of washed red blood cells (RBC´s) in a buffer solution show very little shear thinning which results only from the deformability and orientability of the RBC´s (Maeda et al. 1983). Resuspending them in the plasma protein fibrinogen yields in a much more pronounced shear thinning of the suspension. The invention of the rheoscope in the 1970´s (Golstone et al. 1970) showed that this can be directly related to the formation of clusters of RBC´s at low shear rates. These clusters look similar to stacks of coins that is why they are also called *rouleaux* (fig. 1). *Rouleaux* formation has been studied in vitro frequently by replacing the fibrinogen by the neutral synthetic polymer dextran. Depending on the molecular weight and the concentration, very similar rouleaux and a shear thinning viscosity as in the case of fibrinogen are observed. However, for the case of dextran the dependency of the aggregation process on the polymer concentration is non-monotonic. First, aggregation increases with increasing polymer concentration until a maximum is reached and finally it vanishes again at even higher concentrations. The resulting bell shape has been reported first by Chien et al (Jan and Chien 1973) and was later on, at least qualitatively, often confirmed (Brust et al. 2014). Theoretically, there are two competing models that try to explain the effect of aggregation. One is based on the effect of depletion of the polymers between aggregating RBC´s that introduces an osmotic pressure (Neu and Meiselman 2002, Asakura and Oosawa 1954, Baeumler and Donath 1987, Evans and Needham 1988, Steffen et al. 2013). The other theory is called bridging where it is assumed that the polymers reversibly adsorb on the cell surfaces and form bridges if two cells are in close proximity (Brooks 1973, Chien and Jan 1973, Maeda et al. 1987, Merill et al. 1966). Both the bridging and the depletion theory can explain a non-monotonic dependency of the aggregation on the polymer concentration. For the former, a maximum of adhesion is reached when 50% of the surface is covered with polymers but when the surface coverage reaches 100% with increasing polymer concentrations the cells might repel each other due to steric interaction. For the depletion theory it was proposed that neutral polymers can penetrate the RBC´s glycocalyx in the interfacial region between two adhering cells (Neu and Meiselman, 2002). This should occur when the osmotic pressure and thus the concentration is high enough which finally cancels the depletion effect.

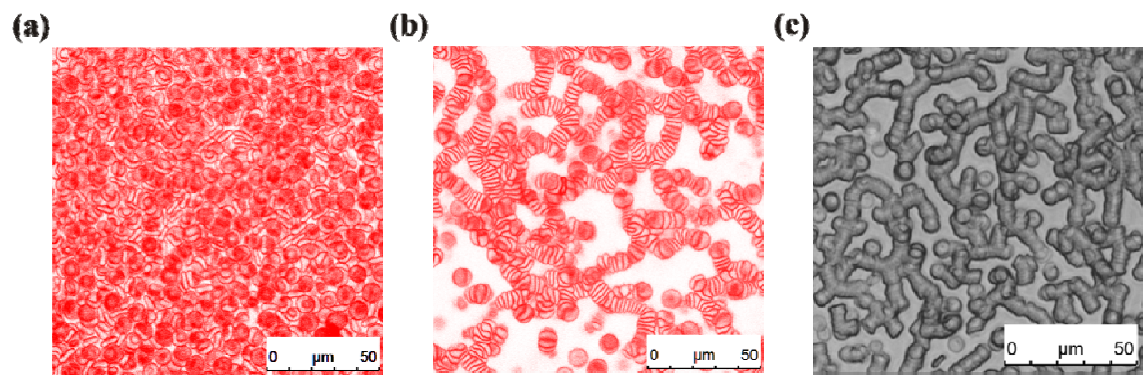

**Fig. 1**: Confocal *microscopic images of RBC`s at 10% hematocrit in the stock solution with (a) 0 mg/ml and (b) 60 mg/ml dextran.(c) Bright light image at 60 mg/ml dextran.*

Adding dextran to the RBC suspensions does not only induce formation of aggregates but it changes also the viscosity of the suspending phase. A common approach to separate the latter effect from the former is to divide the viscosity of the RBC-dextran suspension by the viscosity of the suspending phase. However, the formation of the RBC aggregate is a complex process that involves a rich kinetics. The results will thus not only depend on the chosen protocol but intrinsically on the chosen shear rate, too. Here we will show that measuring the viscosity of a very concentrated suspension of RBC`s of 80% hematocrit, the so called *pellet*, avoids any effects of the kinetics due to the flow and a robust steady state viscosity versus concentration flow curve can be obtained.

## Materials and methods

### Sample preparation

Venous blood of at least three healthy donors was drawn into conventional ethylenediaminetetraacetic acid (EDTA) tubes (S-Monovette; Sarstedt, Nümbrecht, Germany) and washed three times (704 g, 3 min) with phosphate-buffered solution (PBS, Life Technologies, Waltham, MA, USA). This procedure results in a pellet which consists of 80% RBCs and 20% buffer solution, which was verified with a spun hematocrit centrifuge (ZFA (centrifuge), OSC.tec, Erding, Germany) at 10000 and 15000 g.

The dextran solutions were prepared by adding dextran powder (Dextran 70kDa from leuconostoc mesenteroides, Sigma-Aldrich, St Louis, USA) to 1 ml PBS to create a stock solution, with x mg/ml of Dextran (x=0, 20, 40… mg/ml), where the 20% buffer solution in the pellet were taken into account for the determination of the final concentration. To obtain the solutions with RBC-dextran suspension, 4400 µl washed RBC´s (consisting of 80% pure RBC´s and 20% PBS) were added to 3600 µl of the stock solution.

The supernatant and the pellet were prepared after the following protocol: The RBC-dextran suspensions were allowed to sediment in the gravitational field for one hour and were then centrifuged at 704 g for 3 min. The supernatant and the pellet were separated by pipette aspiration. Control experiments with a pure dextran solution showed that this protocol would not cause the dextran to sediment alone.

To ensure that the elevated concentrations of dextran did not affect the RBC´s we checked for an increase of $Ca^{2+}$ concentration in the suspending phase but we could not detect any. An increased level of $Ca^{2+}$ is a typical indicator of a non-physiological or pathological process. Second we checked by visual inspection at large magnification that the RBC´s were still in their physiological discocyte state.

### Rheology

Measurements were done at 23°C with a MCR 702 dual twin rheometer (Anton Paar, Graz, Austria). For the stock solutions, the supernatant and the pellet a cone-plate geometry (CP50/2° diameter: 50mm, angle: 2°) was used. To avoid any effect of sedimentation and to obtain a higher resolution at low shear rates the RBC suspensions at 45% hematocrit with dextran were characterized in a Taylor-Couette geometry (CC20 inner diameter of the outer cylinder: 22mm, diameter of the inner cylinder: 20mm, gap: 1mm, length:50mm ).

The first protocol named "*up*" was performed by first pre-shearing the sample at 100 $s^{-1}$ for one minute followed by a ramp of shear rates from 0.1 $s^{-1}$ to 100 $s^{-1}$ in 16 logarithmical steps with a waiting time of 30 seconds at each step and a time average over 25 seconds. The pre-shearing time of 100 $s^{-1}$ should assure a complete breaking of the rouleaux and we did not find any difference if we pre-sheared the sample for a longer time. The second protocol named "*down*" was performed by first pre-shearing the sample at 100 $s^{-1}$ for one minute followed by a ramp of shear rates from 100 $s^{-1}$ s to 0.01 $s^{-1}$ in 21 logarithmical steps with a waiting time of 30 seconds at each step and a time average over 25 seconds. In general we think that the data below 0.1 $s^{-1}$ s must be taken with care and the error becomes significant. In order to keep the time of our measurement as short as possible to avoid any effect of sedimentation etc. we started the *up*-ramp only at 0.1 but extend until 0.01 at the end of the measurement. All error bars at all measurements show the standard deviation of at least three measurements of three different donors.

These protocols were justified by different tests that showed that sedimentation, or phase separation with the formation of a cell free layer were minimized and the system was in a dynamic steady state, i. e. a constant viscosity was observed for a given shear rate. Therefore we tested the influence of pre-shearing the sample at 100 s$^{-1}$ for 1 minute before each measurement point (Fig. 2a) and compared measurement times of 25 and 50 s for each measurement point. The latter leads to more consistent data at the two lowest shear rates but the other data were equal within experimental resolution. Figure 2a also shows that the dextran solution is a Newtonian fluid with no shear thinning behavior.

In Fig 2b. we show that after 10 seconds a maximum in the viscosity was reached for lowest shear rates for the 60mg/ml dextran-RBC suspension. A slight consecutive decrease in viscosity on a longer time scale might be explained by a slow sedimentation of the RBCs. At higher shear rates the steady state was reached within less than 10 seconds.

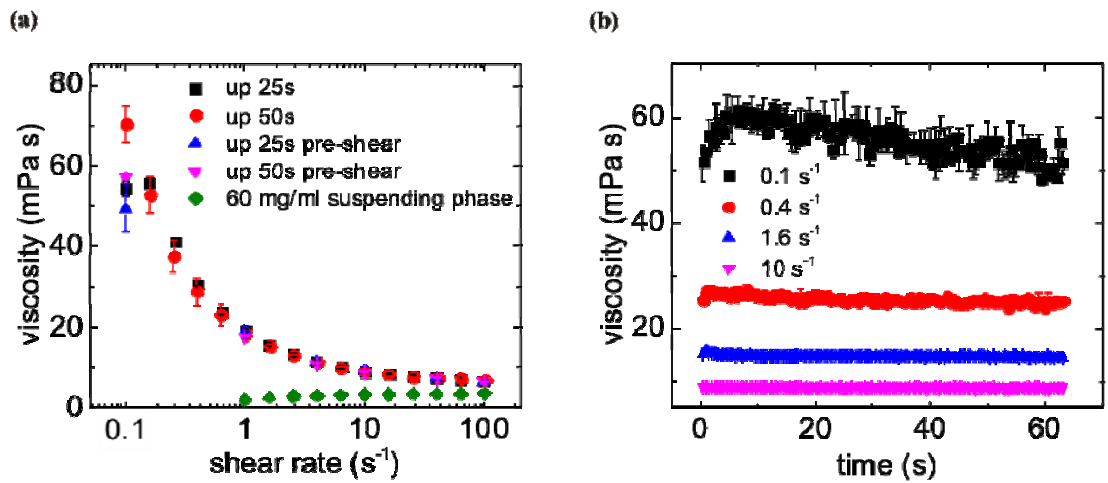

**Fig. 2**: *Test of the up-protocol for a sample with 60 mg/ml dextran. (a) The sample was pre-sheared before integration for 25 (black squares) or 50 (red circles) seconds at the shear rate of the given data point, respectively. In a second run, the sample was pre-sheared at 100 s$^{-1}$ for 25(blue up-triangle) or 50 (pink down-triangle) seconds at each shear rate.(b)The time dependent viscosity after a pre-shear of 100 s$^{-1}$ for 1 minute.*

**Microscopical measurements**

All measurements were performed in an 18 well plate at 23 °C with a hematrocrit of 10 %, which leads to a nearly fully covered well. Cells were labeled with CellMask (Life Technologies, Waltham, MA, USA) by incubating them for 10 min at 23°C. After two times of washing (704 g, 3min) the supernatant was removed and the RBCs were added to the stock solutions. Imaging was performed with a confocal microscope (TSC SP5 II; Leica Microsystems, Mannheim, Germany) with a laser at an excitation wavelength of 633nm. All images were taken after 5±0.5 min as in (Flormann et al. 2015). To check that there was no influence of CellMask on aggregation, several test were performed with unlabeled cells and bright light microscopy without finding any morphological difference in the structure of aggregates.

## Results and discussion

### Microscopical measurements

The confocal images illustrate the effect of rouleaux formation. Figure 1 (a) shows a sample without dextran and the RBC`s are distributed all over on the cover slip at the bottom of the well. Figure 1(b) shows a RBC suspension with a stock solution of 60 mg/ml dextran. *Rouleaux* are clearly to observe, with the RBC´s oriented perpendicular to the horizontal and therefore covering only a fraction of the bottom plate. At higher concentrations, less rouleaux are observed again until they disappear completely at 120 mg/ml. Figure 1(c) shows a measurement at 60 mg/ml without the CellMask in bright light microscopy. Individual cells within a cluster are more difficult to recognize but the result is comparable to the measurements with the confocal microscope.

### Viscosity of the RBC suspensions

Figure 3(a) and (c) show the viscosities of the RBC-dextran suspensions for the *up* and *down* protocol. RBC suspensions without dextran have a rather constant viscosity. The slight increase at low shear rates can be attributed to the resolution limit of the rheometer. With increasing dextran concentrations the viscosities increase as well and the shear thinning becomes more pronounced. The *down* protocol leads to slightly larger viscosities at low shear rates and smaller error bars which allowed, at least in principle, to extend the range of shear rates down to 0.01 s$^{-1}$. However, for a better comparison we did evaluate both data sets only in the range of shear rates from 0.1 to 100 s$^{-1}$. Plotting the viscosities for a given shear rate versus the concentration of the dextran in the stock solution (Fig. 3 (b) and (d)) reveals already a non monotonic behavior at least at very low shear rates. The increase in viscosity with increasing polymer concentration follows both from the formation of aggregates and the increase of the viscosity of the suspending phase (the stock solution) and in the following we will try to separate these two effects.

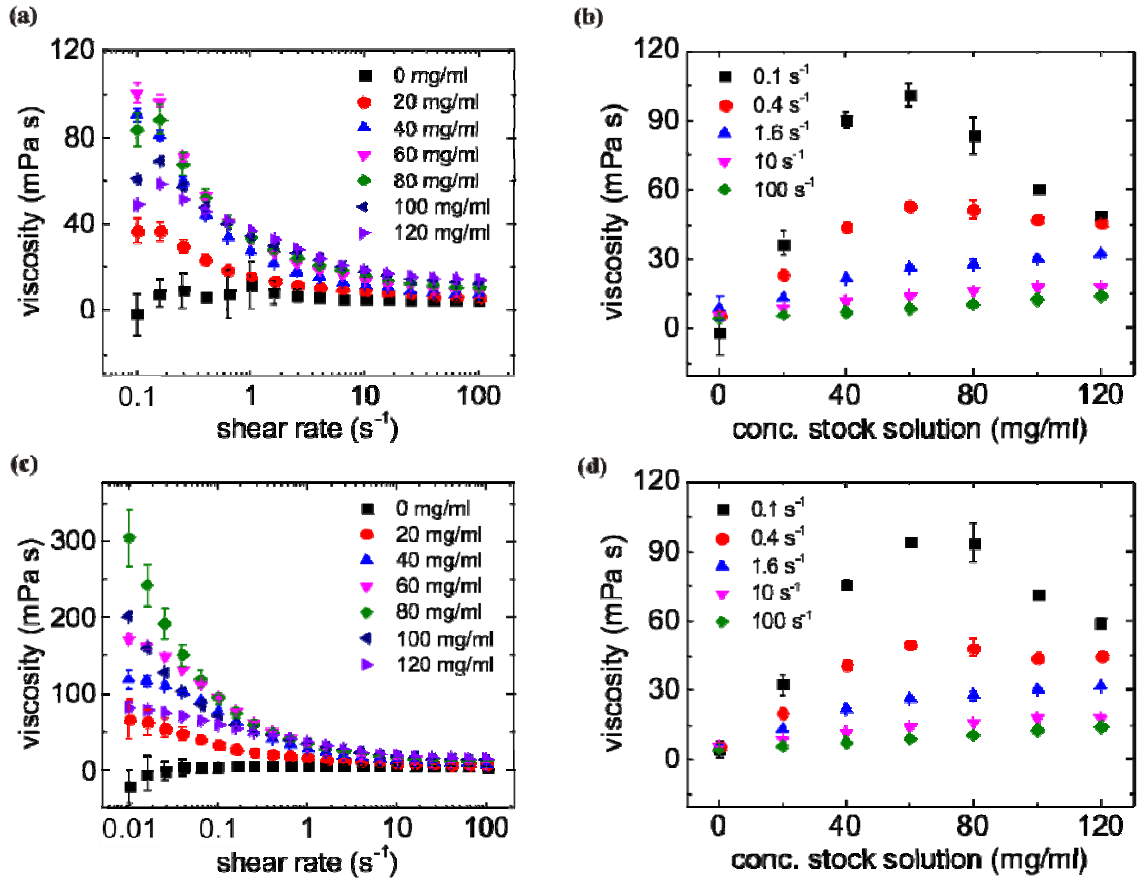

**Fig. 3**: *The viscosity of the RBC-dextran suspension for different dextran concentrations. (a) protocol up (c) protocol down. Respectively, in (b) and (d) the viscosities at a given shear rate as a function of the dextran concentration.*

**Viscosity of the stock solutions and the supernatant**

To separate the effect of aggregation and the increase in viscosity of the suspending phase with increasing polymer concentration one has to determine the viscosity of the suspending phase. At first, this might look as a simple task because the Newtonian viscosity of the dextran solutions can be determined in a straightforward manner (Fig. 4). We found that we could describe the stock solutions viscosity very well with an empirical exponential law

$$\eta(c) = \eta_0 + \eta_1 \cdot e^{b \cdot c}$$

where $\eta$ is the viscosity, $c$ the concentration of dextran $\eta_0$, $\eta_1$ and $b$ fitting constants.

However, it is known that dextran can adsorb to RBC surfaces and the concentration of free dextran in the solution might be less than in the stock solution. The adsorption constant of dextran on RBC has not been consistently determined and we choose to measure the viscosity of the suspending phase in-situ. Therefore the RBC's have been separated again from the dextran solution as described in the materials and methods section and the supernatant viscosity has been determined in the cone-plate geometry (Fig. 4). The viscosity-concentration dependency is qualitatively similar to the stock solution but the viscosities are significantly smaller for the supernatant, meaning that the dextran concentration in the supernatant is reduced and a significant amount of dextran polymers must have been adsorbed on the RBC surfaces.

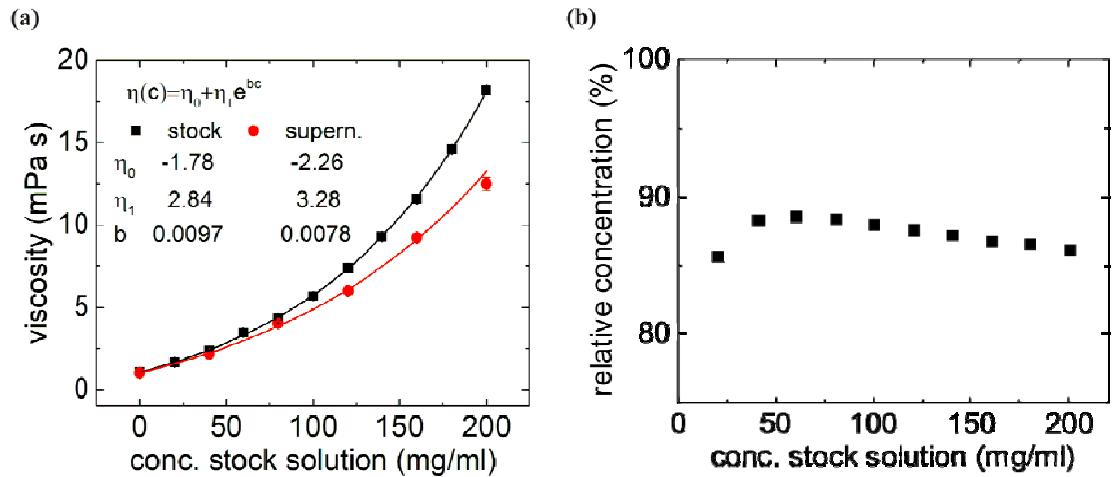

**Fig. 4**: *(a)The viscosities of the stock solutions and of the supernatant. The data are fitted with an empirical exponential law. (b) The relative concentration of macromolecules in the supernatant compared to the stock solution, i.e. the percentage of the non-adsorbed macromolecules.*

**Normalization**

With the viscosity data of the dextran stock solutions we can now normalize the data from the RBC-dextran suspension (Fig. 3 (b) and (d)). Figure 5 shows that the viscosity versus dextran concentration curves have a pronounced bell shape for almost all shear rates. The suspensions have been normalized with both the stock and the supernatant viscosities for both the *up* and *down* protocol. The four curves look qualitatively similar but there are significant differences in the absolute values and in the positions of the maximum of the bell shape. Of course, a normalization with the supernatant viscosity should be the more correct approach but in any case there remains still a dependency of the e.g. position of the maximum of the bell shape on the protocol and especially the shear rates. This means that there is still an underlying kinetic of the formation and breaking of cluster while shearing that might be again affected both by the different adhesion energies and viscosities for the different polymer concentrations. We therefore characterized the flow curve of a pellet of pure RBC´s, i.e. with a very low content of remaining suspending phase after centrifugation in order to minimize the effect of the hydrodynamic coupling.

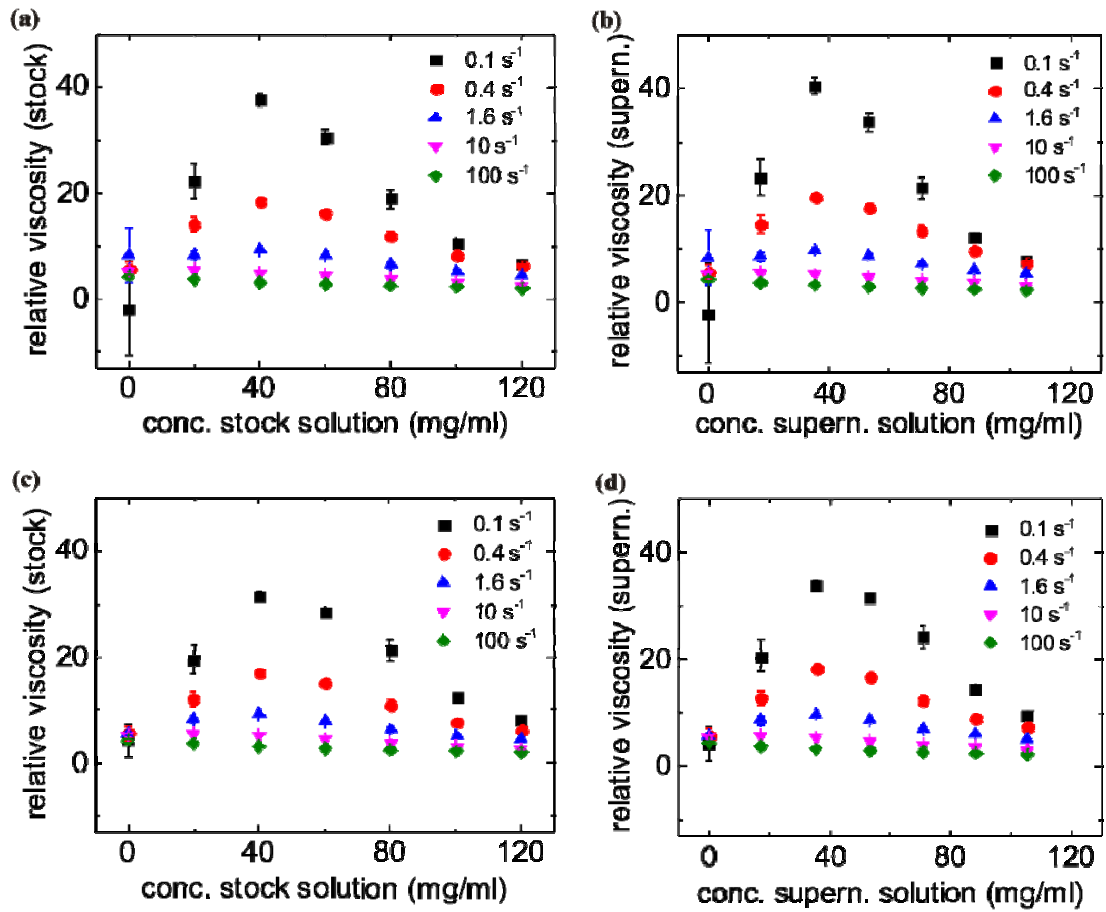

**Fig. 5**: *Relative viscosities of the RBCs - dextran suspensions at different shear rates after normalization with the (a) stock and (b) supernatant viscosities with the up protocol and (c) stock and (d) supernatant with the down protocol.*

**Viscosity of the pellet**

Figure 6 (a) shows the flow curve of the pellet (*up* protocol). Again we observe a shear thinning and the viscosity increases with the dextran concentrations with a maximum at 60 mg/ml and a decrease at higher concentrations. Most importantly, we find that the bell shape curves of the viscosity vs. concentration data differ only in amplitude for the different shear rates but the maximum position of the bell shape is at 60 mg/ml for all shear rates (fig. 6(b)). In order to quantify the difference between the different protocols and samples, the position of the maximum of the bell shaped viscosity vs. concentration curve is shown in Fig. 7. The constancy of the position of the maximum in the pellet measurements indicates that all kinetic effects that are transmitted via the flow field are suppressed due to the very high concentration and in this way we have in especially minimized the effect of the viscosity of the suspending phase. The variations of viscosity versus concentration now result mainly from the dependence of interaction strength on concentration. Still, the viscosity at high dextran concentrations does not go back to exactly the value at zero concentration which probably is a reminiscent of the higher viscosity of the remaining 20% of suspending phase that acts as a lubricant between the cells but apparently this does not affect the position of the maximum of the bell shape viscosity vs. concentration curve.

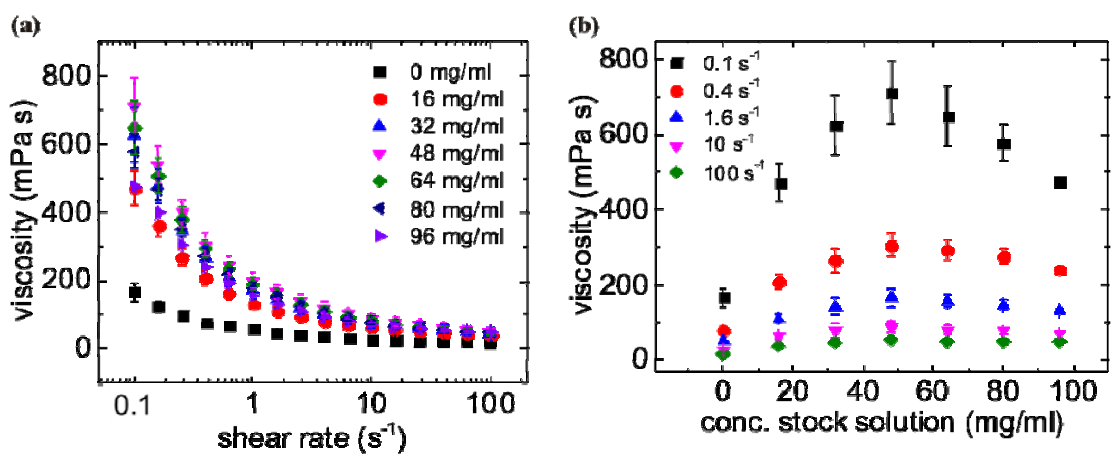

**Fig. 6**: (a) *The flow curve of the pellet (RBCs after sedimentation). (b) The viscosity of the pellet at different shear rates as a function of the dextran concentration.*

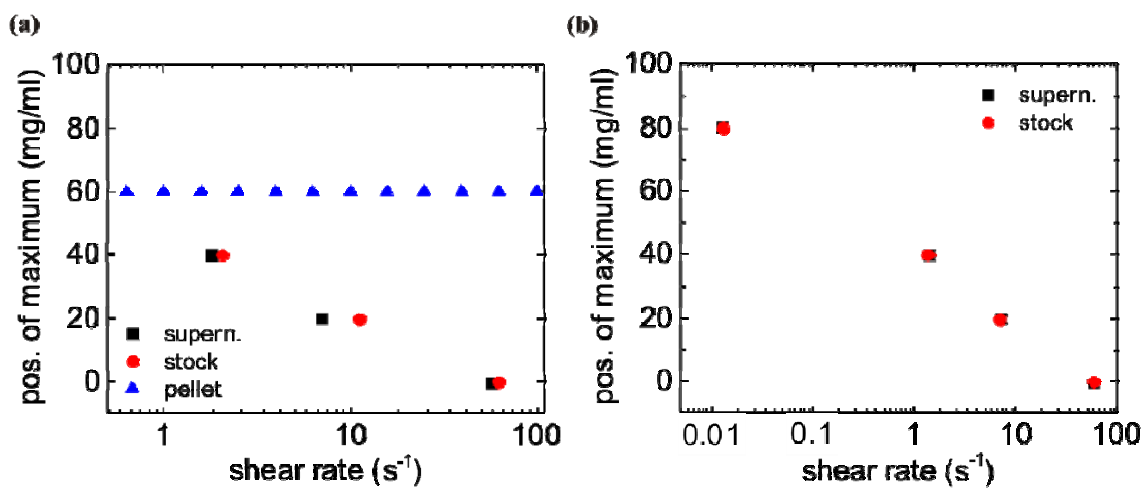

**Fig. 7**: *The position of the maximum of the bell shaped viscosity vs. concentration graphs, for the normalization of the RBC-dextran data with the (a) stock solution and (b) with the supernatant solution and for both protocols up and down. In (a), the constant position of the maximum for the pellet is shown, too.*

## Conclusion

In conclusion we have presented a robust protocol that allows to separate the effect of viscosity increase of the suspending phase and the effect of aggregation in a RBC suspension with different concentrations of dextran. Dextran mimics the physiological situation where the macromolecule fibrinogen causes RBC aggregation. First we have shown that a determination of the relative viscosity of the RBC-dextran suspension is not a straightforward task as dextran adsorbs significantly on the RBC´s and one should carefully consider the supernatant viscosity for a proper normalization. This still does not allow separating the effect of aggregation and increase in viscosity of the suspending phase. We thus have investigated the flow curve of a RBC pellet and indeed found that viscosity vs. concentration curve varies with shear rate only in amplitude while the position of the maximum remains at 60 mg/ml. We should mention that this maximum is in good agreement with our qualitative observations of the microscopic index where we find also the maximum numbers of aggregates at 60 mg/ml.

It is worth mentioning that these results are compatible with what is observed in colloidal systems (Lin et al. 2015). In this case it was found by the shear reversal technique that in a shear thickening suspension of hard micro-spheres at ~45 vol% both hydrodynamic coupling and contact forces contribute to the viscosity. However, it was concluded that only the contact forces cause shear thickening. Here we have a much higher concentration (80 vol%), but no shear thickening which we attribute to the flexibility of the cells that prevents jamming, just like in the case where the RBC´s pass capillaries that are smaller than their diameter without blocking them.

Understanding the rheology and Non-Newtonian fluid mechanics is an important task for the development of predictive tools of e.g. heart and circulatory diseases. The numerical simulation of blood flow has become more realistic nowadays, but the need well controlled quantitative experimental data on the rheology which we hope to provide with this study. Furthermore, there is still search for the ideal plasma expander, and only a sound understanding of the rheological effects that are introduced by the used polymers will allow to understand and to avoid any pathological consequences. Dextran was used for a long term to expand the human plasma and it is still used in veterinary medicine. Due to the fact, that dextran induced substantially problems after applications in humans, it was - even if it still not legally forbidden to use it - substituted by another polysaccharide, namely hydroxyethyl starch (HES) (Lambke and Liljedahl 1976, McCahon and Hardman 2006, Adamik et al. 2015). However, even here serious problems have been identified and therefore it was evaluated as useful for certain patients only who do not have additional diseases (Wiedermann 2014). As dextran is still supposed to be rather inert regarding its biological effect on cells, the pathological effects of dextran are likely purely physical or rheological and we hope that our study will help to better understand the rheological consequences of the use of dextran as a plasma expander.

## Acknowlegments

## Literature

Adamik K, Yozova I, Regenscheit M (2015) Controversies in the use of hydroxyethyl starch solutions in small animals emergency and critical care. Journal of Vetarinary Emergency and Critical care 25 (1): 20-47

Asakura S, Oosawa F (1954) On interaction between two bodies immersed in a solution of macromolecules. J. Chem. Phys. 22: 1255-1256

Baeumler H, Donath E (1987) Does dextran really significantly increase the surface potential of human red blood cells? Stud. Biophys. 129: 113-122

Baskurt O, Neu B, Meiselman H (2012). Red blood cell aggregation. CRC Press.

Brooks D (1973) The effect of neutral polymers on the electrokinetic potential of cells and other charged particles: IV. Electrostatic effects in dextran mediated cellular interaction. J. Colloid Interf. Sci. 43: 714-726

Brust M, Aouane O, Thiébaud M, Flormann D, Verdier C, Kaestner L, Laschke MW, Selmi H, Benyoussef A, Podgorski T, Coupier G, Misbah C, Wagner C (2014) The plasma protein fibrinogen stabilizes clusters of red blood cells in microcapillary flows. Scientific Reports 4: 4348

Chien S, Jan K (1973) Red cell aggregation by macromolecules: role of surface adsoprtion and electrostatic repulsion. J. Supramol. Str. 1: 385-409

Chien S, Usami S, Dellenback R, Gregersen M, Nanninga L, Guest M (1967) Blood viscosity: Influence of erythrocyte aggregation. Science 157: 829-831

Evans E, Needham D (1988) Attraction between lipid bilayer membranes in concentrated solutions of nonadsorbing polymers: comarison of mean-field theory with measurements of adhesion energy. Macromolecules 21: 1822-1831

Flormann D, Kuder E, Lipp P, Wagner C, Kaestner L (2015) Is there a role of C-reactive protein in red blood cell aggregation? Int. Jnl. of Lab. Hem. 37: 474-482

Golstone J, Schoenbein H, Wells R (1970) The rheology of red blood cell aggregates. Microvascular research 2: 273-286

Jan K, Chien S (1973) Role of surface electric charge in red blood cell interaction. The Journal of general physiology 61: 638-654

Lambke L, Liljedahl S (1976) Plasma volume changes after infusion of various plasma expanders. Resuscitation 5: 93-102

Lin N, Guy M, Hermes M, Ness C, Sun J, Poon W, Cohen I (2015) Hydrodynamic and contact contributions to shear thickening in colloidal suspensions. submitted

Maeda N (1983) Alteration of rheological properties of human erythrocytes by crosslinking of membrane proteins. Biochimica et Biophysica Acta 735: 104-112

Maeda N, Seike M, Kume S, Takaku T, Shiga T (1987) Fibrinogen-induced erythrocyte aggregation: erythrocyte-binding site in the fibrinogen molecule. Biochim. Biophys. Acta 904: 81-91

McCahon R, Hardman J (2006) Pharmacology of plasma expanders. Anaesthesia and intensive care medicine 8:2: 79-81

Merill E, Gilland E, Lee T, & Salzmann E (1966) Blood rheology: Effect of fibrinogen deduced by addition. Circ. Res. 18: 437-446

Neu B, Meiselman H (2002) Depletion-mediated red blood cell aggregation in polymer solutions. Biophys. J. 83: 2482-2490

Steffen P, Verdier C, Wagner C (2013) Quantification of depletion-induced adhesion of red blood cells. Phys. Rev. Lett. 110: 018102

Wiedermann C (2014) Reporting bias in trials of volume resuscitation with hydroxyethyl starch. Wien klin. Wochenschau 16: 189-194